# Particle manipulations based on acoustic valley topological rainbow defect-state trapping

**Decai Wu[1], Bowei Wu[1], Tingfeng Ma[1,*], Shuanghuizhi Li[1], Iren Kuznetsova[2], Ilya Nedospasov[2], Boyue Su[1], Teng Wang[1]**

[1] School of Mechanical Engineering and Mechanics, Ningbo University, Ningbo 315211, China

[2] Institute of Radio Engineering and Electronics of RAS, Moscow 125009, Russia

Acoustic microfluidic is an important technology in particle manipulations in biomedical analyses and detections. However, the particle-movement manipulations achieved by the standing surface acoustic wave is suitable for particles in a thin layer of fluids, however it is difficult to manipulate particles in deeper solutions due to the energy loss of surface acoustic waves. The traditional standing bulk wave method can realize the particle manipulation in deep solutions, but it cannot work properly for particle manipulation within a long distance due to the energy loss. In this work, the topological rainbow defect-state trapping is realized, the results show that an effect of point accumulation of acoustic pressure in the waveguide path exists, the position of maximum acoustic pressure can be adjusted flexibly by changing the frequency of the incident acoustic wave, based on which, long-distance movement and capture manipulations of particles in deep solution have been realized. The phenomenon presented in this work can provide a reliable method for manipulations of continuous long-distance particle movement and capture to meet the demand of multiple processing steps in biochemical analyses and detections. The experiment verification results will be presented in the near future.

* matingfeng@nbu.edu.cn

Decai Wu, and Bowei Wu contributed equally to this work.

## I. Introduction

The particle-manipulation technology can realize conveniently position and trajectory controls of particles, enabling fine operations on microscopic objects such as cells and molecules, which has significant application prospects in disease diagnosis, drug delivery, and infection source analysis, et al. [1]. Compared to common manipulation methods like optical tweezers [5], dielectrophoresis [6], thermophoresis [7], and magnetophoresis [8], acoustic tweezers have obvious advantages in terms of penetration, manipulation force, and biocompatibility with biological samples [1]. Many acoustic tweezers based on different acoustic principles have been developed, such as acoustic vortices [9], artificial lens focusing [13], microbubble vortices [14], shape edges vortices [15], structural resonance [19], and ultrasonic transducer arrays [21]. Besides, phononic crystals, as artificially designed periodic structures, have unique capabilities for controlling acoustic and elastic waves [21], thus have broad application prospects in the fields of particle captures, manipulations, and sortings.

In recent years, acoustic topological insulators have attracted much attentions for obvious advantages such as low loss, robustness, and unidirectionality [23]. Acoustic valley topological phases can be constructed by introducing valley degrees of freedom into phononic crystals, which can realize topological transport states at interfaces composed of phononic crystals with different topological phases. Notably, topological interface states have good robustness, and the acoustic and elastic waves protected by topological phases can achieve unidirectional transmission which is resistant to defects and backscattering at the interface [27].

For particle manipulations, by exciting the valley vortex acoustic field of topological insulators, particle patterning can be achieved [29]. Besides, sub-wavelength acoustic manipulation of particles [31], particle manipulations based on higher-order corner states [33] have also been realized. Researchers have also realized manipulation of particle rotation by using the pseudo-spin effect within topological insulators [37], screening of air particles by using phononic crystal waveguides [38], particle manipulation within droplets by using acoustic black holes[39], and concentration and separation of particles within droplets by using scattering standing surface acoustic waves et al. [40].

The particle-movement manipulations achieved by the standing surface acoustic wave is suitable for particles in thin layers of fluids, but it is difficult to manipulate particles in deeper solutions due to the energy loss of the waves. The traditional standing bulk wave method can realize particle manipulations in deep solutions, but it cannot work properly for particle manipulation with a long distance due to the energy loss. In biomedical detection, biological cells in deep solution often need long-distance continuous movement and capture to meet the demand of multiple biochemical processing steps. Therefore it is particularly necessary to explore a method to satisfy the reliable long-distance particle transport in deep solutions. In this work, phononic crystals are used to construct particle transport channels. By realizing topological rainbow defect-state trapping, the position of maximum acoustic pressure can be adjusted flexibly by changing the frequency of the incident acoustic wave, based on which, long-distance movement manipulations and captures of particles in deep solutions have been realized.

## II. Model

A hexagonal honeycomb-lattice phononic crystal is constructed, where the scattering elements are embedded in a liquid matrix, as shown in Fig. 1(a). The lattice constant of the unit cell is $a$, the distance between the center and vertex position is $h$, that between the center and pit position is $b$. Water is selected as the matrix liquid, namely, the mass density is $\rho_w = 1000 \, \text{kg}/m^3$, and the wave speed is $c_w = 1490 \, m/s$. Aluminum is selected as the material of the scatterer, the acoustic impedance of which is $Z_{Al} = \rho_{Al} c_{Al}$, which is 11 times that of water. It can be considered as a rigid wall, meaning that acoustic waves will not transmit through the rigid wall, and all energy will be reflected. FEM numerical simulations of the dispersion relations are carried out by using COMSOL Multiphysics.

By rotating the scatterers within the hexagonal lattice, the spatial inversion symmetry is broken. When $\theta = 0°$ (the scatterers is not rotated), the band structure has a Dirac cone at the K point, as shown by the blue dot in Fig. 1(c). When $\theta = \pm 30°$, the Dirac cone at the K point opens up, forming an acoustic band gap, as indicated by the red dot in Fig. 1(c). From Fig. 1(d), it can be observed that as the angle $\theta$ changes from $-30°$ to $0°$ and then to $30°$, the Dirac cone undergoes a process of "open-close-open". Notably, rotating the scatterers within the unit cell by $-30°$ and $30°$ results in the same band structure, however, the eigenmodes at the upper and lower band-gap edges exhibit a reversal in the pressure field, as shown in Fig. 1(d). This indicates a topological phase transition emerges in the phononic crystal.

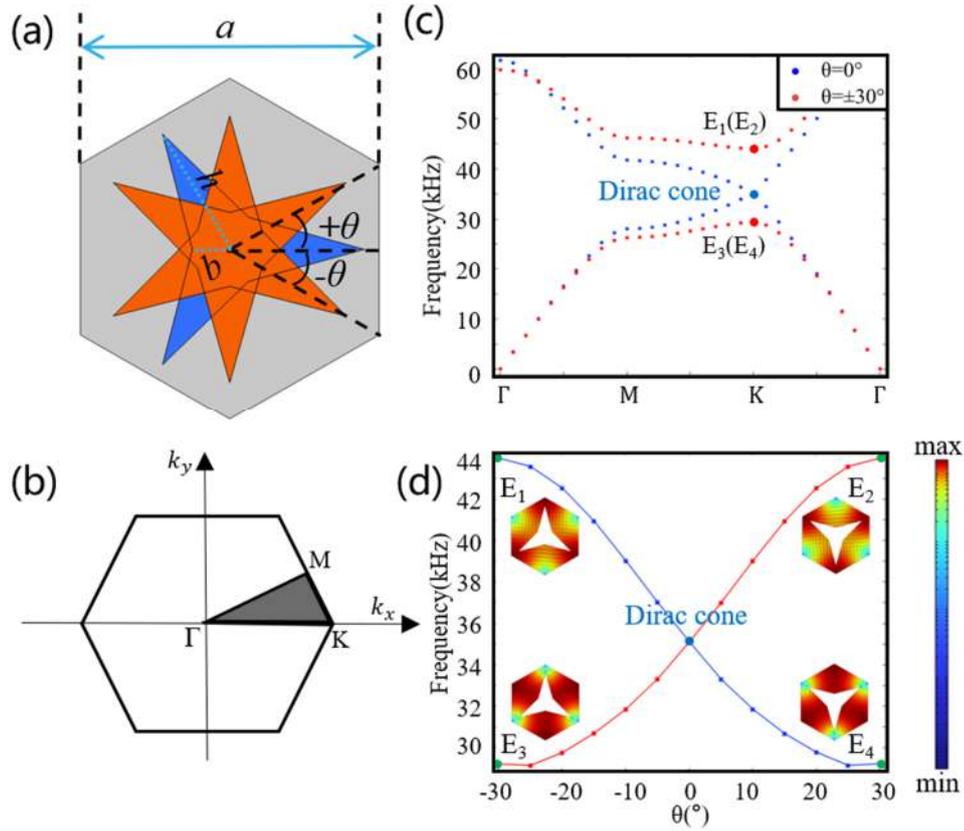

Fig. 1. (a) The unit cell of the phononic cystals with a hexagonal honeycomb lattice, $a$=20mm, $b$=5mm, $h$=9mm. (b) The first Brillouin zone of the unit cell. (c) The dispersion curves of the unit cell. The blue and red dots represent the band structures at rotation angles of 0 and ±30 degrees, respectively. (d) Topological phase inversion of valley pseudo spin states. The inset shows the pressure field distribution of eigenmodes E1-E4，black arrows indicates the direction of the energy

flux.

## III. Topological rainbow defect-sate trapping

Fig. 2(a) shows a supercell consisting of 6 unit cells along the *y*-axis direction. Since the supercell can be considered as a periodically arranged one-dimensional phononic crystal in the *x*-axis direction, Floquet periodic boundary conditions are applied on the left and right boundaries, and plane wave radiation conditions are applied on the top and bottom boundaries to prevent interference caused by the reflection of acoustic waves at the boundaries. The supercell exhibits an A-B structure, where the A-type phononic crystal is composed of unit cells with a scatterer rotation angle of $\theta = 30°$, and the B-type phononic crystal is composed of unit cells with a scatterer rotation angle of $\theta = -30°$. The green line indicates the topological interface. The band structure of the supercell is shown in Fig. 2(b), where the topological transport bands appear in the bandgap between the black bulk bands. The green band represents topological states of acoustic waves at the interface. As shown in Fig.2(b), there is a band gap between the interface's band and the bulk band, which implies that the points of zero group velocity will not be affected by the bulk states. Fig.2(c) shows the acoustic pressure distribution, it can be seen that the energy can be well concentrated at the topological interface. The geometric parameter *b* is gradually increased from 4 mm to 12 mm. Fig. 2 (d) shows the dispersion curves of the supercells with different value of *b*. With the increase of *b*, the topological energy bands of the supercell monotonically shift to higher frequencies.

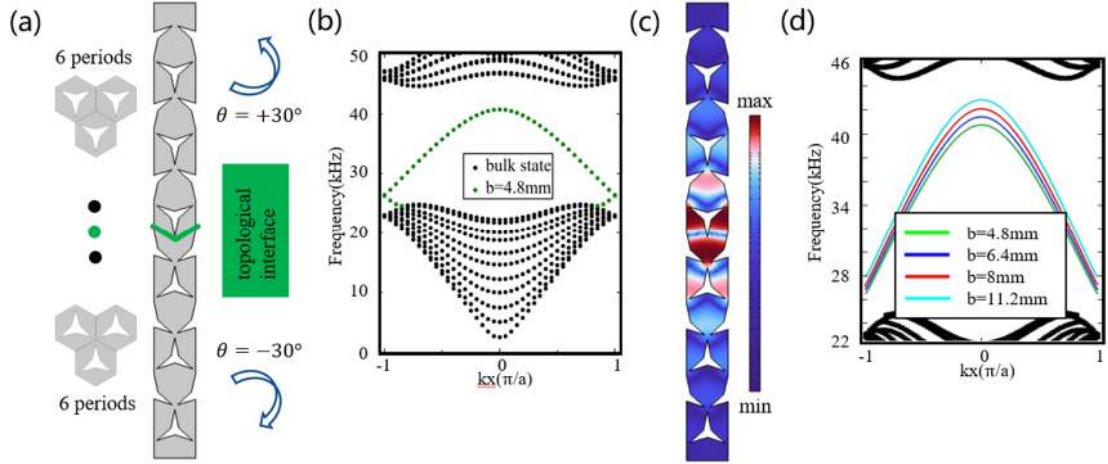

Fig. 2. Dispersion relations and interface mode of the supercell. (a) The supercell structure composed of A-B type phononic crystals; (b) The dispersion relations of the supercell along the Γ-K direction, where black dots represent bulk states, and green dots indicate topological edge states; (c) The acoustic pressure distribution of the topological edge state; (d) The dispersion relations of the supercell with different values of $b$.

The relationship between the group velocity of wave and frequency for different values of $b$ can be reflected by the slopes of the topological band curves. When the group velocity approaches zero, the acoustic waves with the corresponding frequency can no longer propagate forward and are captured at specific locations. As can be seen from Fig. 2(d), with the increase value of $b$, the frequency at which the group velocity is zero gradually increases.

Besides, to verify the rainbow trapping effect of topological insulators, the gradient phononic crystal structure is designed, shown in Fig. 3(a), which is in a rectangular area with a length of 430 mm and a width of 140 mm. There are 20 and 8 unit cells distributed along the $x$-axis and $y$-axis, respectively, where the geometric parameter $b$ of the unit cell increases uniformly and linearly from 4 to 12 mm along the positive $x$-axis. A red dashed line marks the A-B type phononic crystal interface. A plane wave acoustic source with a frequency range of 40 kHz-43 kHz is installed at the center of the right boundary of the structure. In addition, plane wave radiation conditions are applied to the other boundaries of the phononic crystal structure (marked with blue lines) to reduce the interference of reflected waves. As shown in Fig. 3(b), when five incident waves with different frequencies propagate from right to left along the interface, the group velocity gradually decreases to zero and the wave energy is ultimately captured at different locations. Notably, the capture position of low-frequency acoustic waves is closer to the left boundary, while the that of high-frequency acoustic waves is closer to the right boundary. By changing the frequency of the incident waves, the position of acoustic wave capture can be continuously tuned.

When the group velocity approaches zero, the pressure amplitude reaches its maximum. The acoustic wave energy is enhanced by 5-6 times compared to the input. Based on this effect of acoustic energy amplification, even if the power of the incident acoustic wave is not very high, it is sufficient to excite enough acoustic radiation force to drive the movement of particles.

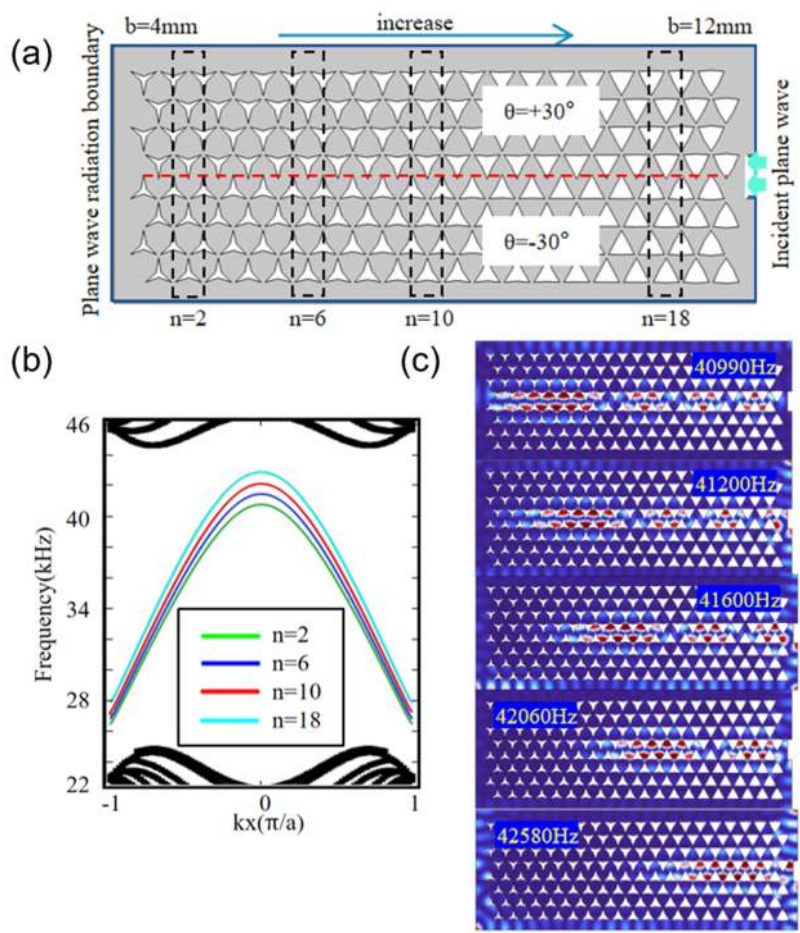

Fig.3 Schematic diagram of the gradient PC structure. (a) The A-B configuration interface is indicated by red dash line, and the geometric parameter $b$ of the scatterers varies linearly along the $x$ direction. (b) Topological energy band diagrams of supercells corresponding to different values

of *n*, the black dot represents bulk states. (c) The distribution of the total acoustic pressure field in the gradient phononic crystal structure under different excitation frequencies.

For the above topological rainbow trapping, the energy concentration spread throughout the part with a positive acoustic wave group velocity, cannot present a phenomenon of point concentration, thus is not adapt to the movement manipulation of particles. Then, topological rainbow with line defect is constructed by removing a layer of scatterers near the interface between two types of phononic crystals, shown in Fig.4(a), forming a central particle transport channel.

For the topological rainbow with a line defect, the dispersion curve of the supercell is simulated, shown in Fig. 4(b). It can be seen that the dispersion curves become flatter compared to that with no defect [43]. The topological rainbow defect-states are located within the yellow square frame. The waveguide effect of topological defect-state rainbow is shown in Fig. 4(c). It can be seen that the energy distribution changes obviously, showing an effect of point accumulation in the whole waveguide path, namely, the energy near the point with a wave group velocity of 0 is stronger than that in other parts of the waveguide. The reason lies in that topological transmission can still be realized when line defects are added to the topological rainbow structure because topological interface states have a certain degree of immunity to defects. However, the line defect causes obvious changes in energy distribution. There exist energy-storage competitions between the points with group velocity 0 and the line defect, and the interfacial energy density is weakened because of a large enough line defect, resulting in that most of the energy concentrates on the area near the point with a wave group velocity of 0. The realization of the energy point-concentration phenomenon provides an important foundation for the captures and continuous long-distance transports of particles.

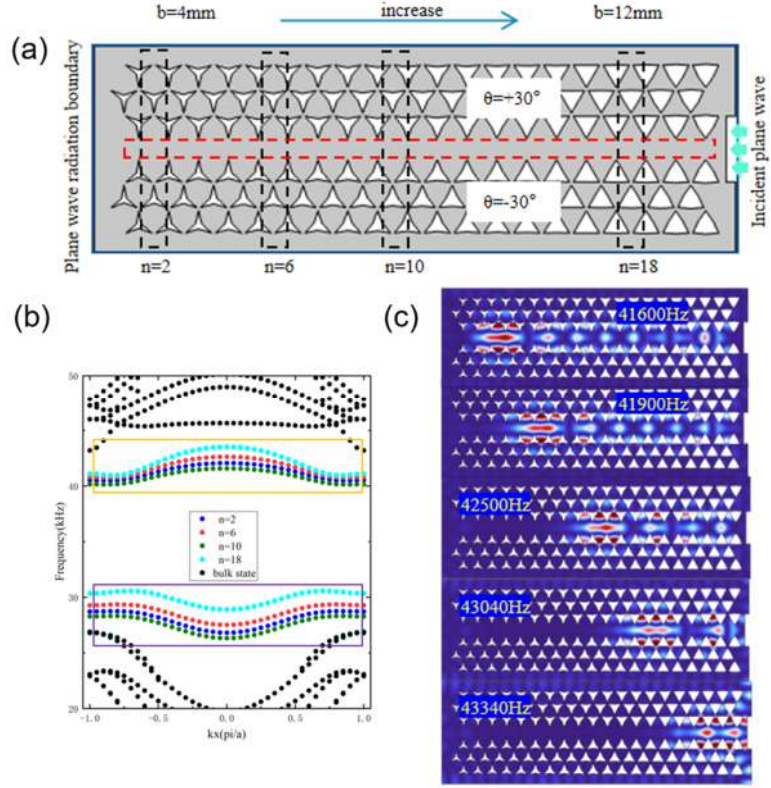

Fig.4 Schematic diagram of topological rainbow defect-state trapping. (a) The A-B type interface is indicated by red dash lines, and the geometric parameter *b* of the scatterer varies linearly along the *x* direction. (b) Dispersion curves of supercells corresponding to different values of *n*, the black dots represent bulk states. The topological rainbow defect-states are located within the yellow square frame. (c) The distribution of the total acoustic pressure field in the topological defect-state rainbow with under different excitation frequencies.

## IV. Particle captures and transports

Based on Gor'kov potential energy theory [44], when the diameter of compressible particles is much smaller than the wavelength of sound ($d_p \ll \lambda$), due to the spatial gradient of the acoustic field, the acoustic radiation force can be expressed as the gradient of a potential function $U_{rad}$. The expression for acoustic radiation force is: $F_{rad} = -\nabla U_{rad}$, where the potential function $U_{rad}$ can be represented by sound pressure and velocity as: $U_{rad} = V_p \left[ f_1 \frac{1}{2\rho c^2} \langle p^2 \rangle - f_2 \frac{3}{4} \rho \langle v^2 \rangle \right]$, where $V_p$ is the volume of the particle, scattering coefficient $f_1 = 1 - \dfrac{\kappa_f}{\kappa_p}$, and $f_2 = \dfrac{2(\rho_p - \rho_f)}{2\rho_p + \rho_f}$ represent monopole and dipole coefficients respectively, $\kappa_f = \rho_f c_f^2$ represents the bulk modulus of the fluid, $\kappa_p = \rho_p \left( c_1^2 - \dfrac{4}{3} c_2^2 \right)$ represents the bulk modulus of the solid, $\rho$ is the density, $\kappa$ is the bulk modulus, subscripts $f$ and $p$ represent the fluid medium and the particle respectively, and $c_1$ and $c_2$ represent the longitudinal and transverse wave speeds of the solid particle material respectively. It is important to note that for this method, based on scattering theory, it is only valid for particles that are much smaller than the wavelength of acoustic waves. When a standing wave field persists in a medium containing particles, if there is a mismatch in acoustic impedance between the fluid and the particle material, the wave will be partially scattered by the particles. The acoustic contrast coefficient depends on the density and compressibility of the particle and its surrounding medium, and its expression is: $\Phi(\kappa, \rho) = \dfrac{5\rho_p - 2\rho_f}{2\rho_p + \rho_f} - \dfrac{\kappa_f}{\kappa_p}$. When the contrast coefficient is positive, the particles move towards the pressure nodes of the standing wave; when the contrast coefficient is negative, the particles move to the high-pressure antinodes.

Polycaprolactone (PCL) is a high-performance polymer with remarkable properties and is widely used in the biomedical field because of several significant advantages [46]. Firstly, PCL exhibits excellent biodegradability and has excellent compatibility with human tissues, allowing cells to grow smoothly on its surface or within it. Besides, PCL does not cause adverse physiological reactions such as blood coagulation when in contact with blood. Crucially, as a hydrophobic material, PCL can effectively regulate the release rate of drugs in the body when used as a drug carrier. However, microspheres made from PCL are physically similar to gels when placed in an aqueous matrix, their acoustic contrast factors satisfy $\Phi(\kappa, \rho) < 0$. This indicates that we can control the position of maximum acoustic pressure by adjusting the incident wave frequency, thereby achieving precise controls over the local position of PCL particles.

COMSOL Multiphysics is used to simulate the manipulation effect of PCL particles. Water is selected as the liquid medium, with a mass density $\rho_w = 1000 \, \text{kg}/m^3$, and a wave speed $c_w = 1490 \, m/s$. The material of the particles is selected as polycaprolactone (PCL), with density of $\rho_p = 1146 \, \text{kg}/m^3$, particle size of $d_p = 500 \, \mu m$, Poisson's ratio of $\mu = 0.4$, and Young's modulus $E = 400 \, MPa$. The compression wave velocity of PCL solid microspheres is $864 \, m/s$, the shear wave velocity is $353 \, m/s$, and the calculated acoustic contrast coefficient is $-1.975$. As shown in Fig.5, the particles are represented by green spheres in the figure. Under the action of acoustic radiation force, the particles move forward and localize to the position of the highest acoustic pressure amplitude in the channel. Then, by monotonically adjusting the frequency of the incident acoustic wave, the movement and capture results of particles are shown in Fig.5. The results indicate that the topological rainbow with defect states can be used to precisely control the forward and backward movement of particles by tuning the operational frequency.

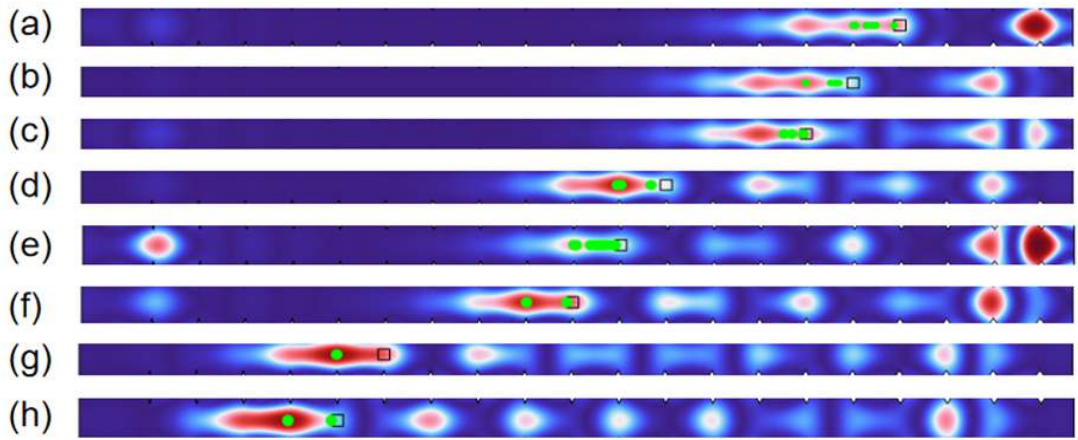

Figure 5: Driving effects of PCL particles in the channel by using the topological rainbow defect-state trapping. The frequencies of incident waves are as respectively: (a) 43080Hz, (b) 42960Hz, (c) 42920Hz, (d) 42500Hz, (e) 42400Hz, (f) 42250Hz, (g) 41710Hz, (h) 41600Hz.

## V. Conclusions

In biomedical detections, biological cells in deep solutions often need long-distance continuous movement and capture to meet the demand of multiple biochemical processing steps. Therefore it is particularly necessary to explore a method to satisfy the reliable long-distance particle transport and capture in deep solutions. In this work, the topological rainbow defect-state trapping is realized, the results show an effect of point accumulation of acoustic pressure in the waveguide path, the position of maximum acoustic pressure can be adjusted conveniently by changing the frequency of the incident acoustic wave, based on which, long-distance movement and capture manipulations of particles in deep solutions have been realized. The phenomenon presented in this work can provide a reliable method for continuous long-distance particle movement and capture manipulations to meet the demand of multiple biochemical processing steps.

It is noted that the experiment verification results will be presented in the near future.

**Acknowledgments**

This work was supported by National Natural Science Foundation of China (No. 12172183)，the Natural Science Foundation of Zhejiang Province (No. LZ24A020001), International Science and Technology Cooperation Project lauched by Science and Technology Bureau of Ningbo City, Zhejiang Province, China (No. 2023H011).